\documentstyle[preprint,aps]{revtex}

\tightenlines

\begin{document}

\draft

\title{
Symmetry Limit Properties of A Priori Mixing Amplitudes for 
Non-leptonic and Weak Radiative Decays of Hyperons
}

\author{
A.~Garc\'{\i}a
}
\address{
Departamento de F\'{\i}sica\\ 
Centro de Investigaci\'on y de Estudios Avanzados del IPN\\
A.P. 14-740. M\'exico, D.F., 07000. MEXICO \\ 
}
\author{ 
R.~Huerta
and
G.~S\'anchez-Col\'on
}
\address{
Departamento de F\'{\i}sica Aplicada\\
Centro de Investigaci\'on y de Estudios Avanzados del IPN. Unidad M\'erida\\ 
A.P. 73, Cordemex. M\'erida, Yucat\'an, 97310. MEXICO\\ 
}

\date{\today}

\maketitle

\begin{abstract}
We show that the so-called parity-conserving amplitudes predicted in the a 
priori mixing scheme for non-leptonic and weak radiative decays of hyperons 
vanish in the strong-flavor symmetry limit.
\end{abstract}


The presence of a priori mixings of strong-flavor and parity eigenstates in 
physical (mass eigenstate) hadrons provides the basis for an explanation of 
the enhancement phenomenon observed in non-leptonic and weak radiative decays 
of hyperons (NLDH and WRDH), as we have shown  elsewhere~\cite{apriori,detailed,universality}.
The transition operators in the amplitudes of these decays are the parity and
flavor conserving strong-interaction Yukawa hamiltonian $H_Y$ and the ordinary 
electromagnetic hamiltonian $H_{em}$.
Although both of them break the strong-flavor symmetries, it is nevertheless
interesting to consider the consequences  of assuming that they do not, i.\ e.,
the consequences on NLDH and WRDH of the symmetry limit behavior of these two
operators.
In this paper we shall demostrate a theorem that says that the so-called
parity-conserving amplitudes in NLDH and WRDH~\cite{ref2} obtained in the a
priori mixing scheme vanish in the strong-flavor symmetry limit.

Let us start with NLDH.
The expressions for the physical hadrons that concern us here in terms of a
priori mixings are~\cite{apriori} 

\[
K^+_{ph} =
K^+_{0p} - \sigma \pi^+_{0p} - \delta' \pi^+_{0s}
+ \cdots
,
\] 

\[
K^0_{ph} = 
K^0_{0p} +
\frac{1}{\sqrt{2}} \sigma \pi^0_{0p} + \frac{1}{\sqrt{2}} \delta' \pi^0_{0s}
+ \cdots
,
\]
 
\begin{equation}
\pi^+_{ph} = 
\pi^+_{0p} + \sigma K^+_{0p} - \delta K^+_{0s}
+ \cdots
,
\label{mph}
\end{equation}
 
\[
\pi^0_{ph} =
\pi^0_{0p} -
\frac{1}{\sqrt{2}} \sigma ( K^0_{0p} + \bar{K}^0_{0p} ) +
\frac{1}{\sqrt{2}} \delta ( K^0_{0s} - \bar{K}^0_{0s} )
+ \cdots
,
\]
 
\[
\pi^-_{ph} =
\pi^-_{0p} + \sigma K^-_{0p} + \delta K^-_{0s}
+ \cdots
,
\]
 
\[
\bar{K}^0_{ph} =
\bar{K}^0_{0p} + \frac{1}{\sqrt{2}} \sigma \pi^0_{0p} -
\frac{1}{\sqrt{2}} \delta'\pi^0_{0s}
+ \cdots
,
\]

\[
K^-_{ph} =
K^-_{0p} - \sigma \pi^-_{0p} + \delta' \pi^-_{0s}
+ \cdots
,
\]

\noindent
and,
 
\[
p_{ph} = 
p_{0s} - \sigma \Sigma^+_{0s} - \delta \Sigma^+_{0p}
+ \cdots
,
\]
 
\[
n_{ph} = 
n_{0s} +
\sigma ( \frac{1}{\sqrt{2}} \Sigma^0_{0s} + \sqrt{\frac{3}{2}} \Lambda_{0s}) +
\delta ( \frac{1}{\sqrt{2}} \Sigma^0_{0p} + \sqrt{\frac{3}{2}} \Lambda_{0p} )
+ \cdots
,
\]
             
\[
\Sigma^+_{ph} =
\Sigma^+_{0s} + \sigma p_{0s} - \delta' p_{0p}
+ \cdots
,
\]
 
\begin{equation}
\Sigma^0_{ph} =
\Sigma^0_{0s} +
\frac{1}{\sqrt{2}} \sigma ( \Xi^0_{0s}- n_{0s} ) +
\frac{1}{\sqrt{2}} \delta \Xi^0_{0p} + \frac{1}{\sqrt{2}} \delta' n_{0p}
+ \cdots
,
\label{bph}
\end{equation}
 
\[
\Sigma^-_{ph} = \Sigma^-_{0s} + \sigma \Xi^-_{0s} + \delta \Xi^-_{0p} 
+ \cdots
,
\]

\[
\Lambda_{ph} = 
\Lambda_{0s} + 
\sqrt{\frac{3}{2}} \sigma ( \Xi^0_{0s}- n_{0s} ) +
\sqrt{\frac{3}{2}} \delta \Xi^0_{0p} + 
\sqrt{\frac{3}{2}} \delta' n_{0p}
+ \cdots
,
\]
 
\[
\Xi^0_{ph} =
\Xi^0_{0s} -
\sigma
( \frac{1}{\sqrt{2}} \Sigma^0_{0s} + \sqrt{\frac{3}{2}} \Lambda_{0s} ) +
\delta'
( \frac{1}{\sqrt{2}} \Sigma^0_{0p} + \sqrt{\frac{3}{2}} \Lambda_{0p} )
+ \cdots
,
\]

\[
\Xi^-_{ph} =
\Xi^-_{0s} - \sigma \Sigma^-_{0s} + \delta' \Sigma^-_{0p}
+ \cdots
.
\]

\noindent
The dots stand for other mixings that will not be relevant here. The 
subindeces naught, $s$, and $p$ mean flavor, positive, and negative parity
eigenstates, respectively.
The amplitudes for NLDH are of the form
$\bar u_{B'} (A+B\gamma_5) u_B$, where $A$ and $B$ are the
so-called parity-violating and parity-conserving amplitudes, respectively.
In what follows, we shall need the expressions for the $B$'s only, they are
given by~\cite{apriori} 

\[
B_1
=
\sigma
(
- \sqrt 3 g_{{}_{p,p\pi^0}} +
g_{{}_{\Lambda,pK^-}} - g_{{}_{\Lambda,\Sigma^+\pi^-}}
)
,\]

\[
B_2
=
-
\frac{1}{\sqrt 2}
\sigma
(
- \sqrt 3 g_{{}_{p,p\pi^0}} +
g_{{}_{\Lambda,pK^-}} - g_{{}_{\Lambda,\Sigma^+\pi^-}}
)
,\]

\[
B_3
=
\sigma
(
\sqrt 2 g_{{}_{\Sigma^0,p K^-}} +
\sqrt{\frac{3}{2}} g_{{}_{\Sigma^+,\Lambda\pi^+}} +
\frac{1}{\sqrt 2}g_{{}_{\Sigma^+,\Sigma^+\pi^0}}
)
,\]

\begin{equation}
B_4
=
\sigma
(
\sqrt 2 g_{{}_{p,p\pi^0}} +
\sqrt{\frac{3}{2}} g_{{}_{\Sigma^+,\Lambda\pi^+}} -
\frac{1}{\sqrt 2}g_{{}_{\Sigma^+,\Sigma^+\pi^0}}
)
,
\label{eq1}
\end{equation}

\[
B_5
=
\sigma
(
g_{{}_{p,p\pi^0}} -
g_{{}_{\Sigma^0,pK^-}} - g_{{}_{\Sigma^+,\Sigma^+\pi^0}}
)
,\]

\[
B_6
=
\sigma
(
- g_{{}_{\Sigma^+,\Lambda\pi^+}} +
g_{{}_{\Xi^-,\Lambda K^-}} + 
\sqrt 3 g_{{}_{\Xi^0,\Xi^0\pi^0}}
)
,\]

\[
B_7
=
\frac{1}{\sqrt 2}
\sigma
(
- g_{{}_{\Sigma^+,\Lambda\pi^+}} +
g_{{}_{\Xi^-,\Lambda K^-}} + 
\sqrt 3 g_{{}_{\Xi^0,\Xi^0\pi^0}}
)
.\]

\noindent
The subindeces $1, \dots, 7$ correspond to
$\Lambda\rightarrow p\pi^-$,
$\Lambda\rightarrow n\pi^0$, 
$\Sigma^-\rightarrow n\pi^-$, 
$\Sigma^+\rightarrow n\pi^+$, 
$\Sigma^+\rightarrow p\pi^0$, 
$\Xi^-\rightarrow \Lambda\pi^-$,
and 
$\Xi^0\rightarrow \Lambda\pi^0$,
respectively.
Here $\sigma$ is the a priori mixing angle that accompanies the positive-parity
eigenstates in the physical hadrons.
Notice the numerical coefficients in these $B$'s.
They are the ones that accompany $\sigma$ in the physical hadrons.
The coupling constants that appear in Eqs.~(\ref{eq1}) are the ordinary
Yukawa couplings observed in strong interactions.
To demostrate the theorem for NLDH we must assume that $H_Y$ is an invariant
operator, an $SU_3$ invariant in the case of Eqs.~(\ref{eq1}).
In this limit the $g$'s are given by~\cite{ref3}

\[
g_{{}_{p,p\pi^0}} = g,\ \ \ \ \ 
g_{{}_{\Sigma^+,\Lambda\pi^+}} = -\frac{2}{\sqrt 3}\alpha g,
\]

\begin{equation}
g_{{}_{\Lambda,\Sigma^+\pi^-}} = -\frac{2}{\sqrt 3}\alpha g,\ \ \ \ \ 
g_{{}_{\Sigma^+,\Sigma^+\pi^0}} = 2(1-\alpha)g,
\label{eq2}
\end{equation}

\[
g_{{}_{\Sigma^0, pK^-}} = (2\alpha-1) g,\ \ \ \ \ 
g_{{}_{\Lambda,pK^-}} = \frac{1}{\sqrt 3}(3-2\alpha)g,
\]

\[
g_{{}_{\Xi^0,\Xi^0\pi^0}} = - (2\alpha-1)g,\ \ \ \ \ 
g_{{}_{\Xi^-,\Lambda K^-}} = \frac{1}{\sqrt 3}(4\alpha-3)g.
\]

\noindent
The connection between $\alpha$ and $g$ and the reduced from factors $F$ and 
$D$ are $\alpha = D/(D+F)$ and $g = D+F$.

When the symmetry limit values of the $g$'s, Eqs.~(\ref{eq2}), are replaced
in Eqs.~(\ref{eq1}), one readily sees that each one of the $B$'s become zero,
i.\ e., $B_1 = B_2 = \cdots = B_7 = 0$, and thus the theorem follows for NLDH.

For WRDH the transition amplitudes are of the form
$\bar u_{B'}(C+D\gamma_5)i\sigma^{\mu\nu} q_\nu u_B \epsilon_\mu$,
where $C$ and $D$ are the so-called parity-conserving and parity-violating
amplitudes, respectively.
$\epsilon_\mu$ is the polarization four-vector of the photon. 
The hadronic parts of the $C$ and $D$ amplitudes are given by~\cite{apriori}

\[
\langle p_{ph} | J^{\mu}_{em} | \Sigma^+_{ph} \rangle =
\bar u_p [ \sigma ( f^{\Sigma^+}_2 - f^p_2 ) +
           (\delta'f^p_2 - \delta f^{\Sigma^+}_2) \gamma^5 ]
i\sigma^{\mu\nu}q_{\nu} u_{\Sigma^+},
\]

\[
\langle \Sigma^-_{ph} | J^{\mu}_{em} | \Xi^-_{ph} \rangle =
\bar u_{\Sigma^-} [ \sigma ( f^{\Xi^-}_2 - f^{\Sigma^-}_2 ) +
                    (\delta' f^{\Sigma^-}_2 - \delta f^{\Xi^-}_2) \gamma^5 ] 
i\sigma^{\mu\nu}q_{\nu} u_{\Xi^-},      
\]

\begin{eqnarray}
\label{eq3}       
\langle n_{ph} | J^{\mu}_{em} | \Lambda_{ph} \rangle 
&=&
\bar u_n 
\left\{ 
\sigma \left[ \sqrt{\frac{3}{2}} ( f^{\Lambda}_2 - f^n_2 ) + 
                   \frac{1}{\sqrt 2} f^{\Sigma^0\Lambda}_2 \right]\right.
\nonumber\\
&&
\left.
+
\left[
\sqrt{\frac{3}{2}} 
(\delta' f^n_2  - \delta f^{\Lambda}_2) 
- 
\delta \frac{1}{\sqrt 2} f^{\Sigma^0\Lambda}_2 
\right]
\gamma^5 
\right\}
i\sigma^{\mu\nu}q_{\nu} u_{\Lambda},
\end{eqnarray}

\begin{eqnarray}
\langle \Lambda_{ph} | J^{\mu}_{em} | \Xi^0_{ph} \rangle 
&=&
\bar u_{\Lambda} 
\left\{ 
\sigma \left[ \sqrt{\frac{3}{2}} ( f^{\Xi^0}_2 - f^{\Lambda}_2 ) -
         \frac{1}{\sqrt 2} f^{\Sigma^0\Lambda}_2 \right]\right.
\nonumber\\
&&
\left.
+                
\left[
\sqrt{\frac{3}{2}} (\delta' f^{\Lambda}_2 - \delta f^{\Xi^0}_2 ) 
+ 
\delta' \frac{1}{\sqrt 2} f^{\Sigma^0\Lambda}_2 
\right] 
\gamma^5 
\right\}
i\sigma^{\mu\nu}q_{\nu} u_{\Xi^0}, 
\nonumber
\end{eqnarray}

\begin{eqnarray}
\langle \Sigma^0_{ph} | J^{\mu}_{em} | \Xi^0_{ph} \rangle 
&=&
\bar u_{\Sigma^0} 
\left\{ 
\sigma \left[ \frac{1}{\sqrt 2} ( f^{\Xi^0}_2 - f^{\Sigma^0}_2 ) - 
         \sqrt{\frac{3}{2}} f^{\Sigma^0\Lambda}_2 \right]\right.
\nonumber\\
&&
\left.
+                
\left[
\frac{1}{\sqrt 2} (\delta' f^{\Sigma^0}_2 - \delta f^{\Xi^0}_2 )               
+ 
\delta' \sqrt{\frac{3}{2}}f^{\Sigma^0\Lambda}_2  
\right]
\gamma^5 
\right\}
i\sigma^{\mu\nu}q_{\nu} u_{\Xi^0}. 
\nonumber
\end{eqnarray}

\noindent
Here $J^\mu_{em}$ is the electromagnetic current operator.
The origin of the numerical coefficients is the same as in Eqs.~(\ref{eq1}).
In the $SU_3$ symmetry limit the anomalous magnetic moments $f_2$ are related
by 

\begin{equation}
f^{\Sigma^+}_2 = f_2^p,\ \ \ \ 
f^{\Xi^-}_2 = f^{\Sigma^-}_2,\ \ \ \ 
f^{\Xi^0}_2 = f^n_2,
\label{eq4}
\end{equation}

\[
f^{\Sigma^0\Lambda}_2 = \frac{\sqrt 3}{2}f^n_2,\ \ \ \ 
f^{\Sigma^0}_2 =  -\frac{1}{2}f^n_2,\ \ \ \ 
f^\Lambda_2 = \frac{1}{2}f^n_2.
\]

When Eqs.~(\ref{eq4}) are replaced into Eqs.~(\ref{eq3}), the theorem follows
for WRDH, namely, $C_1 = C_2 = \cdots = C_5 = 0$.

A few remarks are in order.
Notice that in the above discussions $\sigma \neq 0$ has been maintained.
This leads to another way to put the theorem.
One may restate it by saying: even if a priori mixings of positive-parity but 
flavor-violating eigenstates are allowed in physical hadrons (that is,
$\sigma \neq 0$) the mixing angle $\sigma$ will drop out of the matrix
elements of the Yukawa and electromagnetic hamiltonian that lead to NLDH and
WRDH in the strong flavor symmetry limit.

Concerning the parity-violating amplitudes $A$ and $D$, no equivalent theorem 
seems to exist.
In the flavor symmetry limit, the only possibility to make these amplitudes
vanish is to required that the a priori mixing angles $\delta$ and $\delta'$
be put equal to zero from the outset.
This is analogous to the case of assuming parity conservation; if indeed
parity is a conserved quantum number one has no choice but to enforce
$\delta=\delta'=0$.
There is, however, an exception in WRDH.
It refers to the charge form factors $f_1$. 
These form factors are directly governed by the charge operator, which is a 
combination of generators that is always conserved.
Once the $f_1$'s are identified with the charges of the physical states,
then the matrix elements of WRDH where they (the $f_1$'s) appear vanish,
even if $\delta,\ \delta'\neq 0$.
This is the reason why the $f_1$'s do not appear in the $D$ amplitudes of 
Eqs.~(\ref{eq4}).
One may then conclude that the parts of parity-violating amplitudes 
directly governed by conserved generators of the flavor-symmetry group will 
be zero even if non-zero parity-violating a priori mixings assumed to exist 
in hadrons~\cite{ref4}.

One should contrast the above theorem with the existing theorems for the 
$W_\mu$-mediated NLDH and WRDH, which are refered to as the
Lee-Swift~\cite{ref5} and Hara~\cite{ref6} theorems, respectively.
Both these theorems state that in the flavor symmetry limit it is the
corresponding parity-violating amplitudes, $A$ and $D$ that vanish~\cite{ref7}.
This illustrates how different are the a priori mixings approach and the
ordinary $W_\mu$-mediated approach to NLDH and WRDH.

There is another theorem available, but due to our current inability to 
compute well with QCD, we our unable to establish a direct connection with it. 
This is the so-called Feinberg-Kabir-Weinberg (FKW) theorem~\cite{ref8} or,
more properly, its extension to quarks.
This extension basically says that if one performes rotations to diagonalize
the quark mass matrix to obtain off-shell quarks, then the rotation angles
drop out, and accordingly an absolutely conserved quantum number can be
redefined, say, strangeness.
The rotations considered in this theorem refer to equal parity (positive)
quarks, mixings with opposite parity quarks have not yet been considered.
The reason why the mixing angles drop out can be traced in the quark
lagrangian to the flavor invariance of the QCD part and to the point-like
e.\ m.\ coupling (only $f_1$-type couplings) of the quarks~\cite{ref9}.
In this perspective, the present theorem might be seen to be the analogous of
the FKW theorem, but this time at the hadron level.
One can also see that at this level the equivalent of the FKW theorem will not
eliminate the a priori mixing angles in general, because hadrons have more
interactions than quarks and of those interactions the ones 
that are symmetry-breaking and not directly controlled by some conserved
combination of generators make the mixing angles to give non-zero observable 
contributions.
Eqs.~(\ref{eq1}) and (\ref{eq3}) are examples of this.
In addition, one can see that another reason for the FKW theorem not to
eliminate the a priori mixing angles at the hadron level is that this theorem
is valid for on-shell quarks and that the passage from the quark to the hadron
level involves off-shell quarks, which are also in a highly non-perturbative
regime, inside hadrons. 
It is probably for these reasons that it is extremely difficult to prove the 
FKW at the hadron level, starting from the quark level.
However, it does not seem unreasonable for us to believe that the theorem
we have discussed in this paper is somehow related to the FKW theorem.

The authors wish to acknowledge partial support from CONACyT (M\'exico).


\begin{references}

\bibitem{apriori}
A.~Garc\'{\i}a, R.~Huerta, and G.~S\'anchez-Col\'on,
J. Phys. G: Nucl.Part. Phys. {\bf 24} (1998) 1207.

\bibitem{detailed}
A.~Garc\'{\i}a, R.~Huerta, and G.~S\'anchez-Col\'on,
J. Phys. G: Nucl.Part. Phys. {\bf 25} (1999) 45.

\bibitem{universality}
A.~Garc\'{\i}a, R.~Huerta, and G.~S\'anchez-Col\'on,
J. Phys. G: Nucl.Part. Phys. {\bf 25} (1999) L1.

\bibitem{ref2}
We remind the reader that in the a priori mixing approach to NLDH and WRDH all 
amplitudes are actually parity and flavor conserving.

\bibitem{ref3}
W. M. Gibson and B. R. Pollard,
{\it Symmetry Principles in Elementary Particle Physics}
(Cambridge: Cambridge University Press 1976).

\bibitem{ref4}
Of course, the $f_1$'s also drop out of the $C$ amplitudes, Eqs.~(\ref{eq3}).

\bibitem{ref5}
B.~W.~Lee and A.~R.~Swift, Phys.\ Rev.\ {\bf 136 B} (1964) 228.

\bibitem{ref6}
Y. Hara, Phys.\ Rev.\ Lett.\ {\bf 12} (1964) 378.

\bibitem{ref7}
The theorem of Ref.~\cite{ref5} refers only to the baryon pole contributions to
the parity-violating amplitudes of NLHD.
The theorem of Ref.~\cite{ref6} is limited to 
$\Sigma^+\to p\gamma$ and $\Xi^-\to\Sigma^-\gamma$.

\bibitem{ref8}
G.~Feinberg, P.~Kabir, and S.~Weinberg, Phys. Rev. Lett. {\bf 3}, 527 (1959). 
For its extension to quarks see the review of J.~F.~Donoghue, E.~Golowich, and 
B.~Holstein, Phys.\ Rep.\ {\bf 131} (1986) 319.

\bibitem{ref9}
A memory of the angles remains in the flavor-violating $W_\mu$ interaction, 
leading to the Kobayashi-Maskawa angles.

\end{references}
\end{document}